\documentclass[aps,prl,twocolumn,superscriptaddress,nofootinbib,notitlepage,longbibliography]{revtex4-2}
\usepackage{amsmath,amssymb,amsfonts,bm,color,graphicx,tabularx}

\usepackage[etex=true,export]{adjustbox}
\usepackage{makecell}
\usepackage{multirow}
\usepackage{amsmath}
\usepackage{verbatim}
\usepackage{longtable}
\usepackage[colorlinks=true,citecolor=cyan]{hyperref}
\hypersetup{colorlinks=true,citecolor=cyan,linkcolor=blue,urlcolor=magenta}
\usepackage{tabularx,multirow,array,diagbox}
\usepackage{adjustbox}
\usepackage{supertabular}

\newcommand{\beq}{\begin{equation}}
\newcommand{\eeq}{\end{equation}}
\newcommand{\bd}{{\bm d}}

\newcommand{\bk}{{\bm k}}

\newcommand{\bsigma}{{\bm \sigma}}

\begin{document}
\title{Nodal Topological Superconductivity Driven by Crystalline Antiunitary Symmetry in Altermagnets}

\author{Xiao Xiao}
\affiliation{Department of Physics, Northeastern University, Boston, Massachusetts 02115, USA}
\affiliation{Quantum Materials and Sensing Institute, Northeastern University, Burlington, Massachusetts 01803, USA}

\author{Arun Bansil}
\affiliation{Department of Physics, Northeastern University, Boston, Massachusetts 02115, USA}
\affiliation{Quantum Materials and Sensing Institute, Northeastern University, Burlington, Massachusetts 01803, USA}

\begin{abstract}
Topological superconductivity hosts protected quasiparticles and is central to topological quantum computation, yet its realization in intrinsic materials remains challenging and often relies on engineered platforms. Here we uncover a symmetry-constrained mechanism for nodal topological superconductivity in altermagnets. Focusing on fourfold rotational collinear altermagnets, we show that the native crystalline antiunitary symmetry $\mathcal{T}C_{4z}$ generically forbids pure spin-singlet pairing and selects pairing structures that admit Bogoliubov-de Gennes (BdG) Hamiltonians with emergent chiral symmetries. These symmetries further give rise to robust nodal topological phases over broad parameter regimes, including a nodal-point phase hosting Majorana flat bands (MFBs) and two distinct nodal-loop phases with chiral Majorana edge states. Notably, the nodal structure persists even after spontaneous breaking of the antiunitary symmetry, indicating that the topology originates from symmetry-constrained pairing rather than direct symmetry protection. Finally, we propose tunneling signatures that can distinguish these nodal phases and probe symmetry breaking experimentally.
\end{abstract}

\maketitle

\textit{Introduction.} Symmetry fundamentally constrains superconducting pairing states \cite{Sigrist1991,Mineev1999}, dictating the allowed pairing channels and their topologies. Realizing topological superconductivity therefore requires symmetry settings that enable nontrivial Bogoliubov-de Gennes (BdG) topology \cite{Sato2017}. However, most existing routes rely on strong spin-orbit coupling (SOC) \cite{Alicea2012,Pientka2018,Ren2019}, engineered heterostructures \cite{Lutchyn2018,Lutchyn2010,Oreg2010,Ghorashi2024}, or fine-tuned symmetry breaking \cite{Pientka2013,Nadj_Perge2014}, often requiring delicate tuning and limiting intrinsic material realizations.

Altermagnets (AMs), a recently identified class of magnetic systems \cite{Smejkal2020,Smejkal2022a,Smejkal2022b}, provide a distinct symmetry setting and host a broad range of emergent properties, including topological and correlated phenomena \cite{Ghorashi2024,Smejkal2023,Bhowal2024,McClarty2024,Leeb2024, Yu2025,Li2023,Li2024,Fernandes2024,Antonenko2025,Parshukov2025a}, unconventional transport responses \cite{Ma2021,Cui2023a,Chen2023,Guo2023,Cui2023b,Yang2024,Attias2024, Zhou2024,Syljuasen2025,Zhang2024,Sun2023a,Papaj2023,Beenakker2023, Ouassou2023,Lu2024,Cheng2024,Wei2024,Banerjee2024,Cheng2024a, Chakraborty2025,Sim2025,Froldi2025,Froldi2026,Zyuzin2024,Hu2024}, and spintronic applications \cite{Smejkal2022c,Bai2023,Sun2023b,Zhangr2024,Hodt2024,Chi2024}. Despite a vanishing net magnetization, they exhibit momentum-dependent spin splitting governed by spin-space symmetries tied to the underlying lattice, even in the absence of SOC. Moreover, AMs can host crystalline antiunitary symmetries, formed by combining time-reversal $\mathcal{T}$ with spatial operations, arising from the interplay between spin and lattice degrees of freedom \cite{Xiao2024,Chen2024,Jiang2024}. These unconventional symmetry structures, together with momentum-dependent spin splitting, raise the question of how they constrain superconducting pairing and topology \cite{Mazin2025,Zhu2023,Brekke2023,Carvalho2024,Chakraborty2025a, Feng2025,parshukov2025b,Heung2025,Wu2025,Fukaya2026,zou2026,luo2026}.

Here we uncover a symmetry-driven mechanism for nodal topological superconductivity arising from the crystalline antiunitary symmetry $\mathcal{T}C_{4z}$ in collinear AMs. Group-theoretical analysis reveals that $\mathcal{T}C_{4z}$ admits two irreducible representations (IRRs) for pairing channels: those in the trivial IRR preserve the symmetry, whereas those in the nontrivial IRR spontaneously break it.

\begin{figure}[h]
  \centering
  \includegraphics[width=\columnwidth]{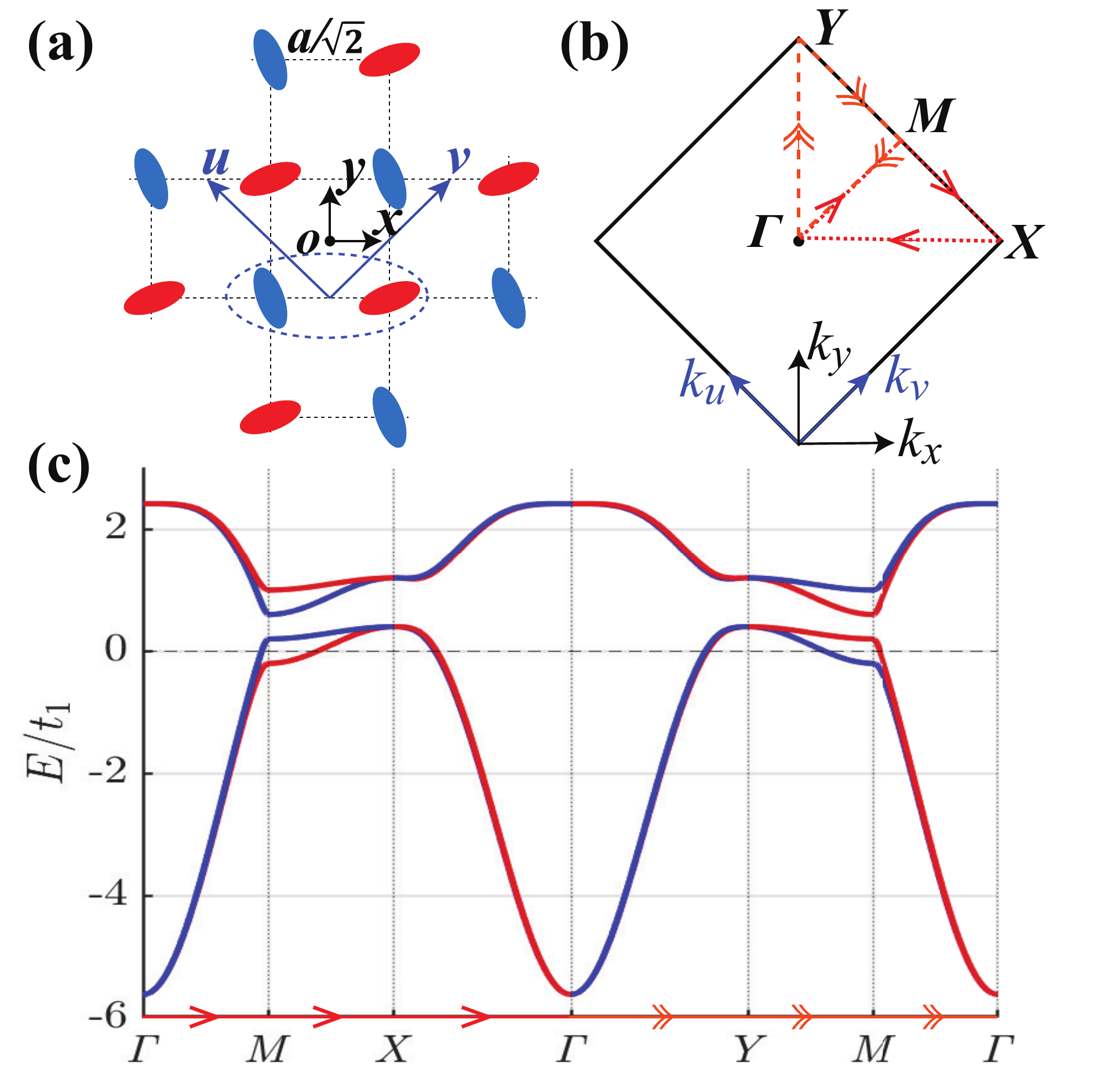}
  \caption{Normal-state properties: (a) lattice structure; (b) Brillouin zone in which a high-symmetry momentum path is shown; (c) Band structure along the path in (b) with chemical potential $\mu=0$. In (c), red (blue) curves denote spin-up (spin-down) bands. The hopping parameters are $t_2=0.3t_1$, $t_3=0.05t_1$, $t_4=0.1t_1$, $t_5=0.02t_1$, and $J=0.4t_1$.}
  \label{fig1}
\end{figure}

We analyze the pairing energetics using linearized gap equations with on-site and nearest-neighbor interactions. A key consequence of symmetry and pairing interactions is the generic mixing between spin-singlet and spin-triplet components along the spin-quantization axis. We find that the leading pairing channels give rise to BdG Hamiltonians with emergent global chiral symmetries, which stabilize nodal topological superconductivity over broad parameter regimes. For moderate interaction strengths, the system hosts either a nodal-point phase characterized by nontrivial winding numbers or a nodal-loop phase protected by $\mathbb{Z}_2$ invariants, depending on the chemical potential. In the nodal-point phase, the chiral symmetry enforces zero-energy boundary states, giving rise to Majorana flat bands (MFBs) \cite{Wong2013,Daido2017}. At stronger interactions, a nodal-loop phase that spontaneously breaks $\mathcal{T}C_{4z}$ can also emerge, with topology likewise characterized by $\mathbb{Z}_2$ invariants. Finally, we demonstrate distinct tunneling signatures that distinguish these nodal phases and diagnose whether the antiunitary symmetry is preserved or spontaneously broken.

\textit{Model of the AM normal state.}
A generic fourfold rotational collinear AM with $\mathcal{T}C_{4z}$ symmetry can be realized on a two-sublattice square lattice [Fig.~\ref{fig1}(a)], with the corresponding Brillouin zone shown in Fig.~\ref{fig1}(b). With Pauli matrices $\tau_i$ and $\sigma_i$ acting on sublattice and spin spaces, the normal-state Hamiltonian, including hopping up to the third-nearest-neighbor sites, is
\beq \label{Hamiltonian}
\mathcal{H}(\bk)=
-c_1(\bk)\tau_x\sigma_0
-c_2(\bk)\tau_0\sigma_0
+c_3(\bk)\tau_z\sigma_0
+J\tau_z\sigma_z,
\eeq
where $c_1=2t_1(\cos\frac{k_x}{\sqrt{2}}+\cos\frac{k_y}{\sqrt{2}})$, $c_2=4t_2\cos\frac{k_x}{\sqrt{2}}\cos\frac{k_y}{\sqrt{2}}+2t_4(\cos\sqrt{2}k_x+\cos\sqrt{2}k_y)$, and $c_3=4t_3\sin\frac{k_x}{\sqrt{2}}\sin\frac{k_y}{\sqrt{2}}+2t_5(\cos\sqrt{2}k_x-\cos\sqrt{2}k_y)$. We set the primitive lattice constant to $a=1$. The $c_3$ term breaks mirror symmetries of the square lattice, leaving $\mathcal{T}C_{4z}$ as the only remaining symmetry. The band structure along the high-symmetry path marked in Fig.~\ref{fig1}(b) is shown in Fig.~\ref{fig1}(c), where the characteristic momentum-dependent spin splitting of altermagnets is clearly visible.

\begin{table}[t]
\begin{tabular}{c|c|c|c} \hline \hline
IRR & $W$ & singlet & triplet \\ [0.5ex]

& & $\psi(\bk)(i\sigma_y)$ 
& $(\bd(\bk)\cdot\bsigma)(i\sigma_y)$ \\

\hline

$\Gamma_1$ 
& $|\mathcal{U}|$ 
& $\psi_{0;on}^{\Gamma_1}
=f_{0;0}^{\Gamma_1}\tau_0
+f_{0;z}^{\Gamma_1}\tau_z$ 
& $d_z^{\Gamma_1}(\bk)
=C_z^{\Gamma_1}(\bk)\tau_y$ \\

& 
& $\psi_0^{\Gamma_1}(\bk)
=C_0^{\Gamma_1}(\bk)\tau_x$ 
& $d_{\pm}^{\Gamma_1}(\bk)
=iS_{\pm}^{\Gamma_1}(\bk)\tau_x$ \\

\hline

$\Gamma_2$ 
& $-|\mathcal{U}|$ 
& $\psi_{0,1}^{\Gamma_2}(\bk)
=S_0^{\Gamma_2}(\bk)(i\tau_y)$ 
& $d_{z,1}^{\Gamma_2}(\bk)
=iS_z^{\Gamma_2}(\bk)\tau_x$ \\

& & 
& $d_{\pm,1}^{\Gamma_2}(\bk)
=C_{\pm}^{\Gamma_2}(\bk)\tau_y$ \\

\hline \hline
\end{tabular}
\caption{Pairing functions forming bases closed under the antiunitary symmetry $\mathcal{T}C_{4z}$. The structure factors are $C_m^{\Gamma}(\bk)= f_{m;x}^{\Gamma}\cos\frac{k_x}{\sqrt{2}}+f_{m;y}^{\Gamma}\cos\frac{k_y}{\sqrt{2}}$ and $S_m^{\Gamma}(\bk)=f_{m;x}^{\Gamma}\sin\frac{k_x}{\sqrt{2}}+f_{m;y}^{\Gamma}\sin\frac{k_y}{\sqrt{2}}$, where $m=0,z,\pm$ label the singlet, $z$-component triplet, and $\pm$-component triplet channels. The relation between $f_{m;x}^{\Gamma}$ and $f_{m;y}^{\Gamma}$ depends on whether $\mathcal{T}C_{4z}$ is preserved or broken. For $\Gamma_2$, the doubled components are omitted for clarity and can be generated by applying $\mathcal{T}C_{4z}$ to $\psi_1(\bk)$, $d_{z,1}(\bk)$, and $d_{\pm,1}(\bk)$.}
\label{table:1}
\end{table}

\textit{Pairing channels constrained by $\mathcal{T} C_{4z}$.}
The allowed superconducting pairing channels are constrained by the symmetry group $\mathcal{G}$ generated by $\mathcal{T}C_{4z}$. This group can be decomposed as $\mathcal{G}=\mathcal{U}\oplus \mathcal{T}C_{4z}\,\mathcal{U}$, where $\mathcal{U}$ is the unitary subgroup generated by $g_1=(\mathcal{T}C_{4z})^2$. Since $\mathcal{U}$ is isomorphic to the cyclic group $C_4=\{e,g_1,g_1^2,g_1^3\}$, it admits four IRRs.

To classify pairing channels according to these IRRs, we construct the projection operator
\beq
P^{\Gamma_n}=\frac{1}{|\mathcal{U}|}\sum_{g_j\in\mathcal{U}}\chi^{\Gamma_n}(g_j)^*\,D(g_j),
\eeq
where $|\mathcal{U}|$ is the order of $\mathcal{U}$, $\Gamma_n$ labels the IRR, $\chi^{\Gamma_n}(g_j)$ is the corresponding character, and $D(g_j)$ is the representation acting on Bloch states. Applying the projectors yields pairing functions for each IRR of $\mathcal{U}$.

In the presence of the antiunitary symmetry, the pairing functions for $\mathcal{G}$ are obtained from those of $\mathcal{U}$ via the Wigner test \cite{Dresselhaus2008},
\beq
W=\sum_{g_j\in\mathcal{U}}\chi^{\Gamma_n}\!\left((\mathcal{T}C_{4z}g_j)^2\right).
\eeq
If $W=|\mathcal{U}|$, the pairing functions of $\mathcal{U}$ directly furnish those of $\mathcal{G}$. If $W=-|\mathcal{U}|$ or $W=0$, the representation is doubled by including the action of $\mathcal{T}C_{4z}$ on the pairing functions of $\mathcal{U}$.

Assuming pairing up to nearest neighbors, a straightforward analysis shows that the pairing functions in two of the four IRRs vanish, while those in the remaining two IRRs are summarized in Table~\ref{table:1}. Since these pairing functions form a complete basis, the gap function can be expanded accordingly and recast in terms of an orthonormal basis $\{\Psi_{m;a}^{\Gamma}(\bk)\}$ (see Supplemental Material for details \cite{supporting}) as
\beq
\Delta(\bk) = D_0 \left[\eta_{0;0}^{\Gamma_1} \Psi_{0;0}^{\Gamma_1} + \eta_{0;z}^{\Gamma_1} \Psi_{0;z}^{\Gamma_1} + \sideset{}{'}\sum_{\Gamma,m,a} \eta_{m;a}^{\Gamma} \Psi_{m;a}^{\Gamma}(\bk)\right],
\eeq
where the first two on-site pairing terms are written separately from the primed nearest-neighbor summation; $\Gamma\in\{\Gamma_1,\Gamma_2\}$ labels the IRR, $m\in\{0,z,\pm\}$ the pairing channel, and $a\in\{x,y\}$ the component index; the coefficients $\eta_{m;a}^{\Gamma}$ denote the corresponding order parameters, and $D_0$, taken as $0.1t_1$ throughout this work, sets the overall superconducting gap scale.

\begin{figure}[h!]
  \centering
  \includegraphics[width=\columnwidth]{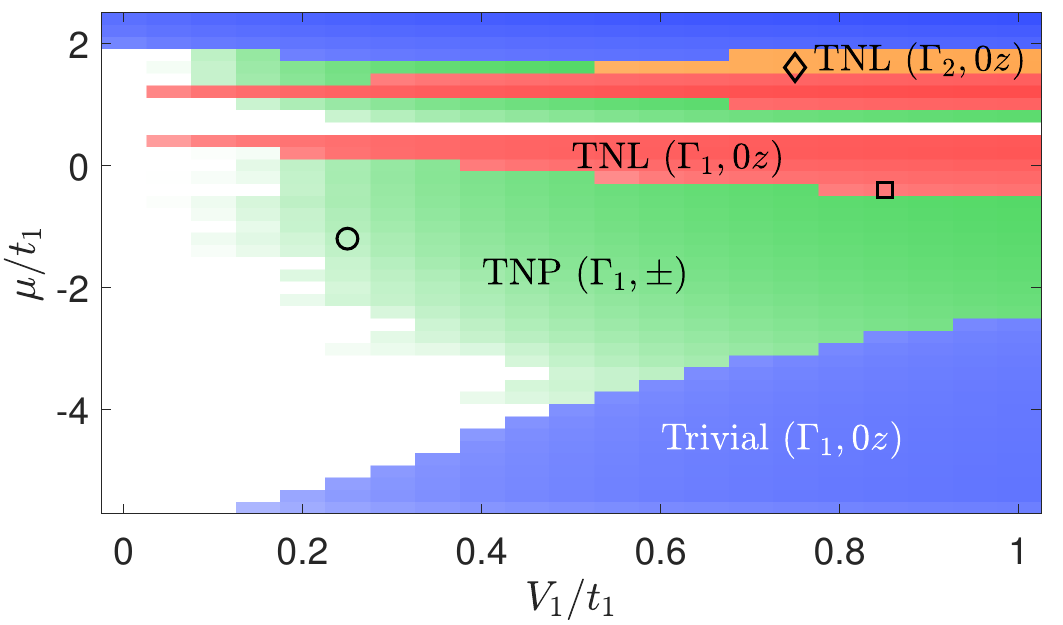}
  \caption{Phase diagram of the leading superconducting instability for the model with $\mathcal{T}C_{4z}$ as a function of $\mu$ and $V_1$, with $V_0/t_1=0.5$. Colors denote different leading pairing phases, while brightness encodes the magnitude of $T_c$. The lowest visible value is $k_B T_c/t_1 = 10^{-6}$.}
  \label{fig2}
\end{figure}

\textit{Mixing of singlet and $z$-component triplet.}
To identify generic energetic constraints, we expand the interaction in the pairing basis as \cite{Goryo2012,Yuan2014}:
\begin{align}
V(\bk,\bk') = &-V_{0} \left[\Psi_{0,0}^{\Gamma_1}(\bk)\, \Psi_{0,0}^{\Gamma_1}(\bk')^\dag + \Psi_{0,z}^{\Gamma_1}(\bk)\, \Psi_{0,z}^{\Gamma_1}(\bk')^\dag \right] \nonumber \\
&-V_{1} \sideset{}{'}\sum_{\Gamma,m,a} \Psi_{m;a}^{\Gamma}(\bk)\, \Psi_{m;a}^{\Gamma}(\bk')^\dag,
\end{align}
where $V_0$ and $V_1$ denote the coupling strengths for the on-site and nearest-neighbor channels, respectively. Near the superconducting onset, the gap function is small, and the self-consistent gap equation can be linearized to yield (see Supplemental Material for details \cite{supporting}):
\begin{align} \label{LGE}
\eta_{m;a}^{\Gamma} &= V_{m,a}^{\Gamma}\, k_B T \sum_{\Gamma',m',a'} \eta_{m';a'}^{\Gamma'} \nonumber \\
& \times \sum_{\bk',n} \mathrm{Tr} \left[\Psi^{\Gamma}_{m;a}(\bk')^\dag \mathcal{G}_{\bk'}(i\omega_n)\, \Psi^{\Gamma'}_{m';a'}(\bk')\, \bar{\mathcal{G}}_{-\bk'}(i\omega_n)\right],
\end{align}
where $V_{m,a}^{\Gamma}=V_0$ for $(\Gamma,m,a)=(\Gamma_1,0,0)$ or $(\Gamma_1,0,z)$, and $V_{m,a}^{\Gamma}=V_1$ otherwise. Here $\mathcal{G}_{\bk'}$ and $\bar{\mathcal{G}}_{-\bk'}$ denote the normal-state Matsubara Green’s functions for electrons and holes, respectively. 

For a collinear altermagnetic normal state, the spin sector of the Green’s functions can be separated and has the form $P_s=(\sigma_0+s\sigma_z)/2$ with $s=\pm1$. The spin structures of the pairing channels are $i\sigma_y$ for the singlet ($m=0$) and $\sigma_z(i\sigma_y)$ for the $z$-component of the triplet ($m=z$). Consequently, 
$\mathrm{Tr}\!\left[(i\sigma_y)P_s\,\sigma_z(i\sigma_y)P_{s'}\right]$
is nonzero whenever $s\neq s'$. This implies that, in collinear altermagnets, the singlet channel generically mixes with the $z$-component of the triplet channel, constituting the first key result of this work.

\textit{Phase diagram.}
From Eq.~(\ref{LGE}), we determine the phase diagram of the leading pairing instabilities, characterized by $T_c$, in the $(\mu, V_0, V_1)$ parameter space. A representative phase diagram for $V_0/t_1=0.5$ is shown in Fig.~\ref{fig2}. At the bottom and top of the bands (blue regions), the singlet-triplet mixed $0z$ channel in $\Gamma_1$ dominates. Owing to dominant on-site pairing components, these phases are trivial and fully gapped. However, between the two trivial regions, three distinct topological nodal phases emerge over a broad parameter range, which we discuss below.

\begin{figure}[t!]
  \centering
  \includegraphics[width=\columnwidth]{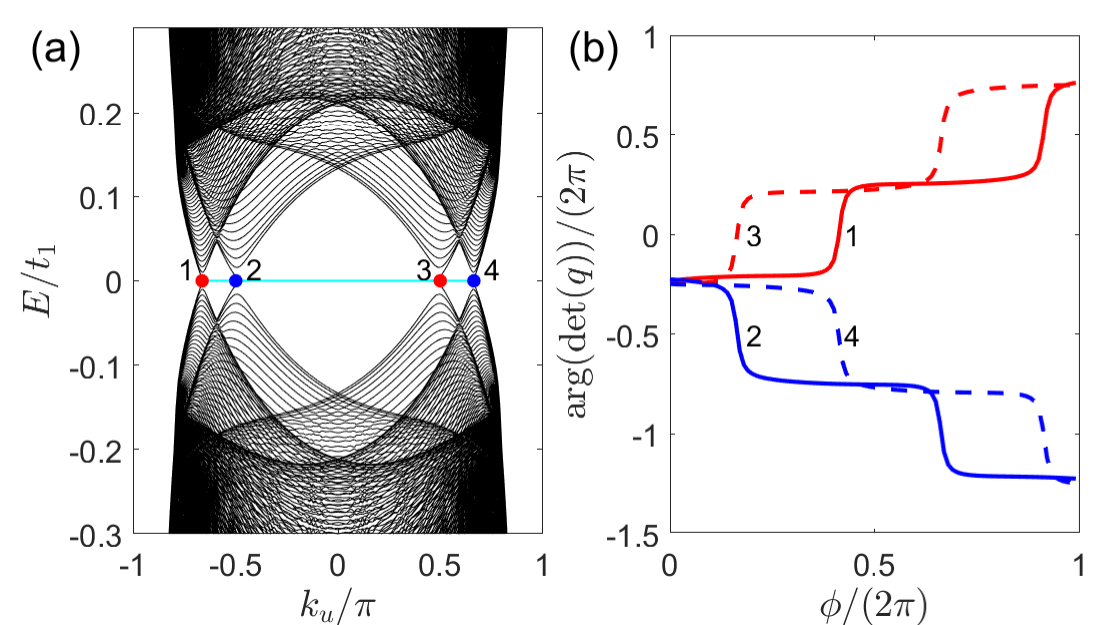}
  \caption{Topological nodal-point superconducting phase: (a) Low-energy spectrum of a finite-size system with $L=500a$ as a function of $k_u$, with edge states highlighted in cyan. The four dots denote the projections of the bulk nodal points onto the $k_u$ axis; red (blue) dots correspond to nodes with winding number $\nu=1$ ($\nu=-1$). (b) Phase winding of $\det[q(\bk)]$ around the four nodal points, evaluated along closed loops encircling each.}
  \label{fig3}
\end{figure}

\textit{Topological nodal-point superconductivity and MFBs.}
In the green regions of Fig.~\ref{fig2}, the $+$ and $-$ triplet pairing channels in $\Gamma_{1}$ are degenerate at the leading instability. The $\mathcal{T}C_{4z}$ symmetry further constrains the order parameters as $\eta_{+,x}^{\Gamma_1} = \pm i (\eta_{-,y}^{\Gamma_1})^*$ and $\eta_{+,y}^{\Gamma_1} = \mp i (\eta_{-,x}^{\Gamma_1})^*$. As a result, the BdG Hamiltonian exhibits a global chiral symmetry $\mathcal{C} = \kappa_y \otimes \tau_0 \otimes P_{s=1} + \kappa_x \otimes \tau_0 \otimes P_{s=-1}$, where $\kappa_i$ are Pauli matrices in Nambu space and $P_{s=\pm1}$ are the spin projectors defined below Eq.~(\ref{LGE}). In the chiral basis of $\mathcal{C}$, the Hamiltonian takes an off-diagonal form with blocks $q(\bk)$ and $q^\dag(\bk)$.

Since $\det(q(\bk))$ is generally complex, the nodal condition requires both $\Re[\det(q(\bk))]=0$ and $\Im[\det(q(\bk))]=0$. In two dimensions, these two constraints generically yield isolated solutions, corresponding to geometrically stable nodal points. The complex $\det(q(\bk))$ exhibits a nontrivial winding around these nodes, endowing them with topological protection characterized by a winding number $\nu$. These features are confirmed in Fig.~\ref{fig3} at a representative parameter point (denoted by the circle in Fig.~\ref{fig2}). Crucially, in this phase $\mathcal{C}$ is preserved locally at each $\bk$, rather than only globally through the BdG relation between $\bk$ and $-\bk$, thereby pinning the boundary spectrum to zero energy and giving rise to MFBs.

\begin{figure}[h!]
  \centering
  \includegraphics[width=\columnwidth]{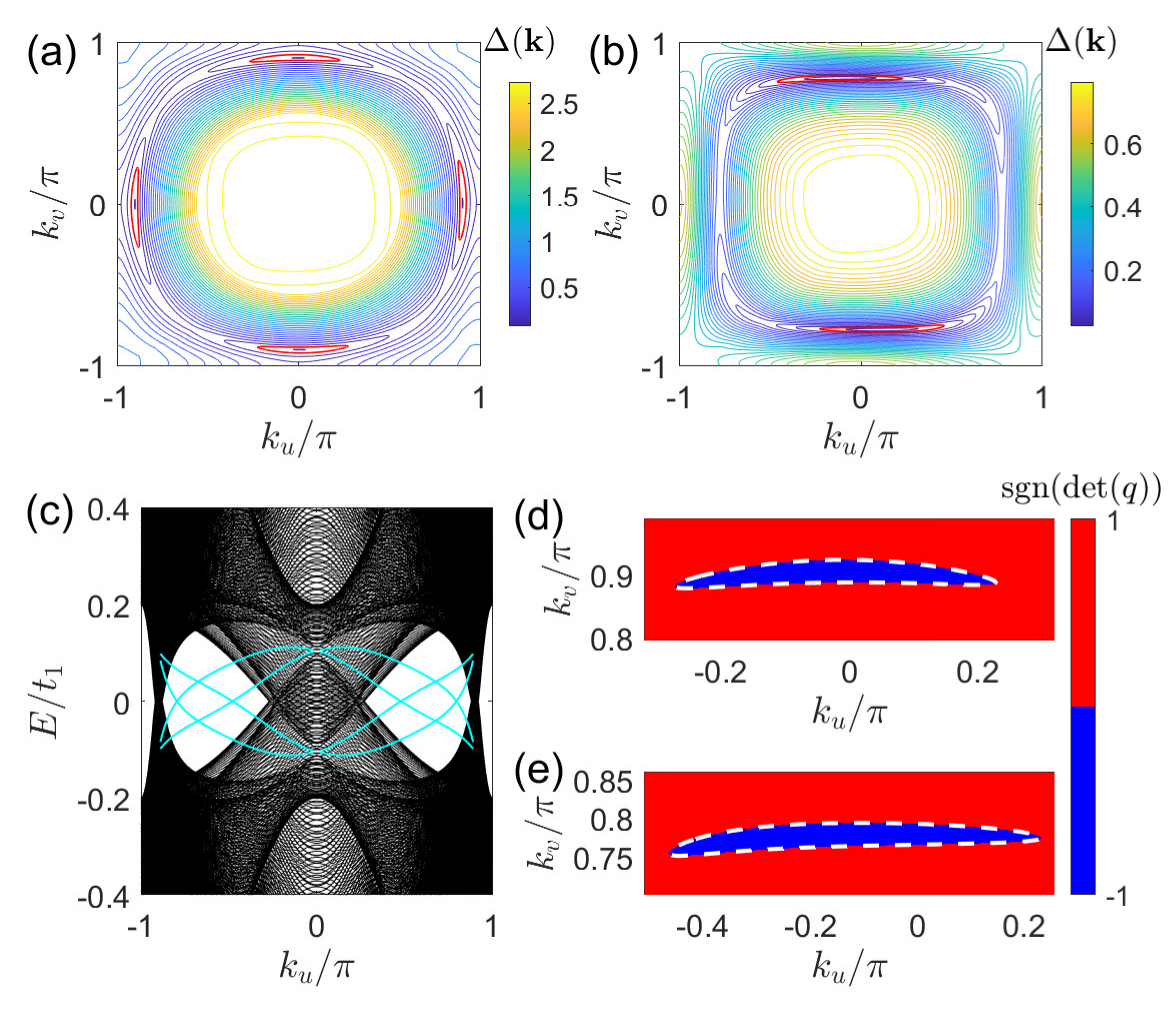}
  \caption{Topological nodal-loop superconducting phases: (a,b) Gap distributions in the Brillouin zone for the $0z$ channels in $\Gamma_1$ and $\Gamma_2$, respectively, with nodal loops highlighted in red. (c) Low-energy spectrum of a finite-size system ($L=500a$) as a function of $k_u$ in the $\Gamma_1$ TNL phase, with edge states highlighted in cyan. (d,e) $\mathbb{Z}_2$ topology of the nodal loops, revealed by the sign of $\det(q)$ along representative loops (white dashed curves) for the TNL phases in $\Gamma_1$ and $\Gamma_2$, respectively.}
  \label{fig4}
\end{figure}

\textit{Topological nodal-loop superconductivity.}
In the red regions of Fig.~\ref{fig2}, Eq.~(\ref{LGE}) selects a state with vanishing $\eta_{0;0}^{\Gamma_1}$ and finite $\eta_{0;z}^{\Gamma_1}$. The $\mathcal{T}C_{4z}$ symmetry further constrains the order parameters to be purely imaginary, such that $\eta_{0;a}^{\Gamma_1}=\pm (\eta_{0;a}^{\Gamma_1})^*$ and $\eta_{z;a}^{\Gamma_1}=\pm (\eta_{z;a}^{\Gamma_1})^*$. The resulting BdG Hamiltonian possesses a global chiral symmetry generated by $\kappa_x$. In addition, an emergent antiunitary symmetry $\mathcal{A}=\kappa_z K$, with $K$ denoting complex conjugation, enforces the off-diagonal block $q(\bk)$ to be real. Consequently, the nodal structure is determined solely by $\Re\det[q(\bk)]=0$. In two dimensions, this single constraint generically gives rise to geometrically stable nodal loops. Their topology is characterized by a $\mathbb{Z}_2$ invariant given by $\mathrm{sgn}[\det(q)]$, which supports topologically protected chiral edge states.

In the orange region of Fig.~\ref{fig2}, the leading instability corresponds to the $0z$ channel in $\Gamma_2$. In this case, $\mathcal{T}C_{4z}$ enforces the corresponding order parameters to vanish, so the selection of this phase by Eq.~(\ref{LGE}) spontaneously breaks $\mathcal{T}C_{4z}$. From the resulting order parameters, we again identify a chiral symmetry generated by $\kappa_y$, together with an emergent antiunitary symmetry $\mathcal{A}=(\kappa_z \otimes u)K$ with $u=\mathrm{diag}(1,-1,1,-1)$, which similarly constrains $q(\bk)$ to be real. Consequently, the system also realizes a topological nodal-loop superconducting phase, but with $\mathcal{T}C_{4z}$ spontaneously broken.

The numerical results in Fig.~\ref{fig4} at two representative parameter points (denoted by square and diamond symbols in Fig.~\ref{fig2}) corroborate this analysis. In the $\Gamma_1$ phase, the preservation of $\mathcal{T}C_{4z}$ leads to four symmetry-related nodal loops [Fig.~\ref{fig4}(a)], whereas in the $\Gamma_2$ phase the spontaneous breaking of $\mathcal{T}C_{4z}$ reduces the nodal structure to two loops [Fig.~\ref{fig4}(b)], explicitly breaking fourfold rotational symmetry.

\begin{figure}[h!]
  \centering
  \includegraphics[width=0.95\columnwidth]{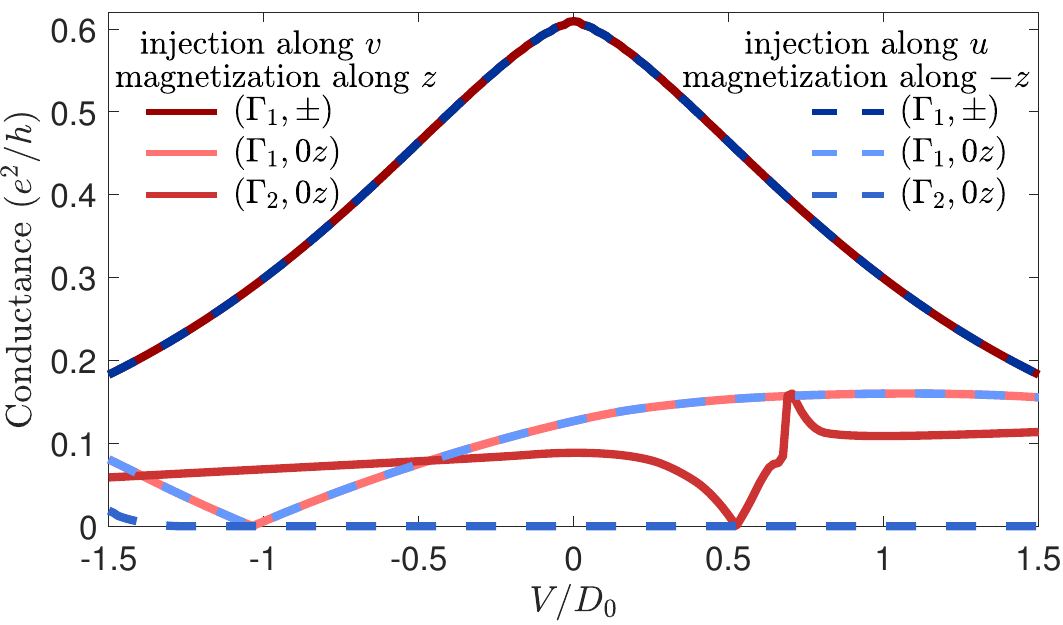}
  \caption{Tunneling spectra of planar junctions between an altermagnetic superconductor and a half-metallic lead. Solid (dashed) curves correspond to injection along the $v$ ($u$) direction with the lead spin polarization aligned along $z$ ($-z$). The two configurations are related by $\mathcal{T}C_{4z}$. Colors distinguish the pairing channels $(\Gamma_1,\pm)$, $(\Gamma_1,0z)$, and $(\Gamma_2,0z)$.}
  \label{fig5}
\end{figure}

\textit{Tunneling fingerprints of the nodal phases.}
These three topological nodal phases can be distinguished via tunneling spectroscopy in two planar junctions with half-metallic leads \cite{Keizer2006} related by $\mathcal{T}C_{4z}$. One configuration has the junction along $v$ with the magnetization in the lead along $z$, while $\mathcal{T}C_{4z}$ maps it to a second configuration along $u$ with the magnetization along $-z$ (Fig.~\ref{fig5}).

In the TNP phase, MFBs induce resonant equal-spin Andreev reflection \cite{He2014}, yielding a pronounced zero-bias conductance peak. In contrast, for the TNL phases in the $0z$ channel, Andreev reflection is forbidden for $z$-polarized half-metal leads, so low-energy transport is dominated by quasiparticle tunneling and the conductance is strongly suppressed.

Both the TNP phase and the TNL phase in $\Gamma_1$ preserve $\mathcal{T}C_{4z}$, leading to identical spectra for the two configurations. By contrast, in the $\Gamma_2$ TNL phase, the spontaneous breaking of $\mathcal{T}C_{4z}$ results in distinct tunneling spectra.

\textit{Influence of SOC.}
We have focused on the spin-conserving limit in which the nodal phases are protected by chiral and/or real symmetries. In the presence of SOC, these protecting symmetries might be broken \cite{Roig2025}. In these cases, the nodal structures become fully gapped but may still realize distinct topological superconducting states \cite{Sato2017}. In particular, upon gapping, the system may have a nonzero Chern number and realize a chiral topological superconducting phase \cite{Zhu2023,Read2000}. Moreover, in the present model, the interplay between SOC-induced mass terms and the underlying rotational structure naturally leads to edge-dependent mass configurations, in which the mass acquires different signs on edges related by rotation. This provides a natural route to a higher-order topological superconducting phase, characterized by domain walls at the corners and the emergence of Majorana zero modes \cite{Zhu2023}. The absence of mirror and inversion symmetries further removes constraints that would otherwise enforce cancellation of topological contributions \cite{Nagaosa2010,Fu2011}, making the emergence of chiral or higher-order topology a natural consequence once SOC gaps the nodal phases\footnote {A detailed investigation of these possibilities would be interesting, but it is beyond the scope of this work.}.

\textit{Material realizations.}
A relevant materials context for the present model can be identified from two complementary directions. First, recent classifications of two-dimensional altermagnets \cite{Zeng2024} have identified candidate materials with spin groups containing $4'$ (equivalent to $\mathcal{T}C_{4z}$), such as V$_2$Se$_2$O \cite{Ma2021}, V$_2$Te$_2$O \cite{Cui2023a}, and Cr$_2$O$_2$ \cite{Chen2023,Guo2023}. These systems already realize the key symmetry underlying our model, although they typically possess additional crystalline symmetries. In practice, such additional symmetries may obscure the minimal symmetry setting considered here, but can be selectively reduced, for example, by substrate effects \cite{Novoselov2016} or strain \cite{Ando2015}, thereby bringing the system closer to the $\mathcal{T}C_{4z}$ limit. 

A complementary route is provided by quasi-two-dimensional organic charge-transfer salts such as $\kappa$-Cl \cite{Yu2025}. These systems realize correlated electrons on an anisotropic triangular lattice characterized by hopping amplitudes $t$ and $t'$, and therefore do not possess the $\mathcal{T}C_{4z}$ assumed here. However, their electronic structure might be continuously tuned through pressure, chemical substitution, or molecular engineering \cite{McKenzie1997,Miyagawa2004}, which modify the ratio $t'/t$. In the limit $t'/t \to 1$, the lattice approaches a square geometry, and the system moves toward the $\mathcal{T}C_{4z}$ symmetry regime. 

Our work therefore isolates the essential role of $\mathcal{T}C_{4z}$ and explores a reduced symmetry setting that may be approached either by symmetry reduction in $4'$-symmetric materials or by lattice tuning in organic systems.

\textit{Conclusions.}
We have demonstrated that in a collinear altermagnetic system with $\mathcal{T}C_{4z}$ symmetry, the crystalline antiunitary symmetry, in conjunction with the altermagnetic spin structure, plays a central role in stabilizing distinct nodal topological superconducting phases. Importantly, these phases emerge naturally from symmetry and are robust over broad parameter regimes, enhancing their experimental accessibility. We further show that they exhibit distinct and experimentally accessible signatures in tunneling spectroscopy, providing a direct route for their detection. Our results establish a symmetry-based framework for realizing topological superconductivity in altermagnetic systems and open a pathway toward robust topological superconductivity beyond conventional symmetry settings.

\textit{Acknowledgment.}
This work was supported by the National Science Foundation through the Expand-QISE award NSF-OMA-2329067 and benefited from the resources of Northeastern University’s Advanced Scientific Computation Center, the Discovery Cluster, the Massachusetts Technology Collaborative award MTC-22032, and the Quantum Materials and Sensing Institute.

\end{document}